\documentclass{article} 
\usepackage{iclr2024_conference,times}

\usepackage{amsmath}
\usepackage{amsfonts}
\usepackage{bm}

\def\eqref#1{equation~\ref{#1}}

\def\1{\bm{1}}

\DeclareMathAlphabet{\mathsfit}{\encodingdefault}{\sfdefault}{m}{sl}
\SetMathAlphabet{\mathsfit}{bold}{\encodingdefault}{\sfdefault}{bx}{n}

\usepackage{hyperref}
\usepackage{url}
\usepackage{multirow}
\usepackage{booktabs}
\usepackage{amsmath}
\usepackage{graphicx}
\usepackage{algorithm}
\usepackage{algpseudocode}
\usepackage{svg}
\usepackage[T1]{fontenc}

\title{Protein-ligand binding representation learning from fine-grained interactions}

\author{Shikun Feng$^1$ \  Minghao Li$^1$ \  Yinjun Jia$^2$ \  Weiying Ma$^1$ \ Yanyan Lan$^{1\dag}$\\
$^1$Institute for AI Industry Research, Tsinghua University\\
$^2$School of Life Sciences, Tsinghua University
}

\iclrfinalcopy 
\begin{document}
\maketitle

\renewcommand{\thefootnote}{}
\footnotetext{$^\dag$Correspondence to \texttt{lanyanyan@air.tsinghua.edu.cn}}
\begin{abstract}

The binding between proteins and ligands plays a crucial role in the realm of drug discovery. Previous deep learning approaches have shown promising results over traditional computationally intensive methods, but resulting in poor generalization due to limited supervised data. In this paper, we propose to learn protein-ligand binding representation in a self-supervised learning manner. Different from existing pre-training approaches which treat proteins and ligands individually, we emphasize to discern the intricate binding patterns from fine-grained interactions. Specifically, this self-supervised learning problem is formulated as a prediction of the conclusive binding complex structure given a pocket and ligand with a Transformer based interaction module, which naturally emulates the binding process. To ensure the representation of rich binding information, we introduce two pre-training tasks, i.e.~atomic pairwise distance map prediction and mask ligand reconstruction, which comprehensively model the fine-grained interactions from both structure and feature space. Extensive experiments have demonstrated the superiority of our method across various binding tasks, including protein-ligand affinity prediction, virtual screening and protein-ligand docking.

\end{abstract}

\section{Introduction}

Understanding the interaction between proteins and ligands is a crucial task in drug discovery, which involves predicting whether the proteins and ligands can bind together or determining the binding affinity and pose of a protein-ligand pair. Deep learning methodologies~\citep{ozturk2018deepdta,abbasi2020deepcda,monteiro2022dtitr,wallach2015atomnet,ragoza2017protein,li2021structure} have become prominent contenders for this direction due to recent and rapid advancements in machine learning. However, the performance of these data-driven methods heavily relies on limited training data and may be susceptible to noise introduced by experimental errors. Therefore, the overall generalizability of these supervised methods is constrained~\citep{shen2021beware}.

Inspired by the remarkable success of self-supervised learning in computer vision and natural language processing, recent works aim to apply it to protein-ligand interactions by utilizing large amount of unlabeled data. The majority of pre-training approaches available today, including ELECTRA-DTA~\citep{wang2022electra}, SMT-DTA~\citep{pei2022smt} and Uni-Mol~\citep{zhou2023unimol}, rely on a two-tower architecture with individual molecular and protein encoders for pre-training. A simple interaction module is then introduced to fine-tune the encoders for downstream binding related tasks. 


However, the binding mechanism of protein-ligand complex is exceedingly intricate, involving a broad range of non-covalent interactions between inter-molecular atom pairs, including  $\pi$-stacking, $\pi$-cation, salt bridge, water bridge, hydrogen bond, hydrophobic interaction and halogen bond\citep{de2017systematic}. Previous studies~\citep{adhav2023realm,ding2022observing} have shown the crucial role of these interactions in determining binding affinity and docking conformation. Furthermore, certain supervised learning methods, for example MONN~\citep{li2020monn}, OnionNet~\citep{zheng2019onionnet} and PIGNet~\citep{moon2022pignet}, have demonstrated notable performance improvement by explicitly incorporating these inter-molecular atom-wise interactions in their methodologies. 

Clearly, the current self-supervised learning methods focus on enhancing the representation of individual molecule or protein, but the interaction module trained in the later fine-tuning stage falls short in capturing these highly intricate interaction patterns. Therefore, how to develop an interaction-aware representation that directly benefits downstream protein-ligand interaction-related tasks remains an unresolved challenge. To our best understanding, CoSP~\citep{gao2022cosp} represents a significant step forward in this direction by leveraging contrastive learning to obtain pocket and ligand representations. While the contrastive learning approach does have the capability to align positive ligand and pocket pairs, it fails to adequately capture the inter-molecular atomic interactions through the global matching manner.




To address this issue, we propose to learn protein-ligand binding representations from fine-grained interactions, named BindNet. Specifically, our self-supervised learning problem involves predicting the conclusive binding complex structure given a primary pocket and ligand, which is in line with the protein-ligand interaction process. To emphasize learning fundamental interactions, we employ a specific Transformer-based interaction module that utilizes individual pocket and ligand encoders in the modeling process. To ensure that the model learns interaction-aware protein and ligand representations, we use two distinct strategies in the pre-training process. The first pre-training objective is Atomic pairwise Distance map Prediction (ADP), where interaction distance map between atoms in the protein and ligand is employed to provide detailed supervision signals regarding their interactions. The other pre-training objective is Mask Ligand Reconstruction (MLR), in which the ligand representation extracted by a 3D pre-trained encoder is masked for reconstruction. By employing feature space masking and reconstruction instead of simply token or atom type masking, the model is more likely to capture richer semantic information, such as chemical and shape information, during the pre-training process, as has been demonstrated in prior works in the fields of computer vision and natural language processing~\citep{baevski2023efficient,assran2023self}.


The primary contribution of this paper can be summarized in four distinct aspects. Firstly, the problem of self-supervised learning of protein-ligand binding representations has been formalized as the prediction of the final complex structure given the primary pocket and ligand structure, which naturally mimic the binding process. Secondly, a new architecture has been designed to incorporate a Transformer-based interaction module on protein and ligand encoders, emphasizing the encoding of intricate interaction representations rather than individual protein and ligand representations. Thirdly, two novel pre-training objectives have been proposed to ensure learning of complex binding patterns for the interaction module. Lastly, extensive experiments have been conducted on a wide range of binding-related tasks, including predicting protein-ligand affinity, virtual screening, and protein-ligand docking, all of which demonstrate the promising results achieved by BindNet.


\section{Related work}

Several pre-training methods have been proposed for proteins and ligands representation learning. DeepAffinity~\citep{karimi2019deepaffinity} utilizes a RNN-based architecture to conduct unsupervised learning based on compound SMILES and protein SPS. Both SMT-DTA~\citep{pei2022smt} and ELECTRA-DTA~\citep{wang2022electra} employ Masked Language Modeling (MLM) to train a molecule encoder based on SMILES and a protein encoder based on amino acid sequences. The distinction between these two methods is that SMT-DTA trains the MLM and the affinity prediction task cocurrently in a multi-task fashion, whereas ELECTRA-DTA utilizes a pre-training and fine-tuning approach. 

Aside from sequence-based methods, there are efforts to incorporate the 3D structure of ligands and pockets for modeling. Uni-Mol, for instance, employs denoising and MLM strategies to train a 3D-based molecular encoder and pocket encoder, which are then fine-tuned for prediction. While CoSP~\citep{gao2022cosp} leverages a contrastive learning framework on pocket-ligand pairs within unlabeled complex data, to pre-train a dual-branch encoder for pockets and ligands.

\section{BindNet}
\label{gen_inst}
This section presents our novel pre-training framework, BindNet.

\vspace{-10pt}
\subsection{Problem Formalization and Model Architecture}
\vspace{-5pt}
In order to capture the complicated protein-ligand interaction patterns, we have formulated the pre-training problem as directly predicting the final complex structure given the structures of both the pocket and ligand. In this formulation, the provided protein structure remains the same as that in the complex, while the given ligand structure represents its primary state, rather than its torsion state after binding. This formulation is more appropriate than using the torsion state as the input structure because it is commonly acknowledged that proteins tend to be relatively rigid, whereas molecules are generally more flexible in the binding process~\citep{stark2022equibind, corso2022diffdock}. 

Specifically, a structured chemical compound is denoted as $\mathcal{G}=(\mathcal{V}, \mathcal{X})$, where $v_i \in \mathcal{V}$ represents the atom type of node $i$, and $x_i \in \mathcal{X}$ represents the 3D position of the i-th atom. Then the full atom-level representation of a protein-ligand complex is denoted as $\mathcal{G}_C=(\mathcal{V}_C, \mathcal{X}_C)$. The pocket and ligand parts are denoted separately as $\mathcal{G}_P=(\mathcal{V}_P, \mathcal{X}_P)$ and $\mathcal{G}_L=(\mathcal{V}_L, \mathcal{X}_L)$, respectively. To ensure the primary molecular structure is used, RDKit is utilized to generate a stable conformation based on the ligand's chemical information. In case of any failures, we propose perturbing the torsion state to approximate the primary structure by introducing Gaussian noise to the dihedral angles or the coordinates of the molecular conformation. The resulted ligand structure is $\mathcal{G}_{\hat{L}}=(\mathcal{V}_{\hat{L}}, \mathcal{X}_{\hat{L}})$.

\begin{figure}[t]
    \vskip -0.4in
    \label{figure:method}
    \begin{center}
    \centerline{\includegraphics[width=1\textwidth]{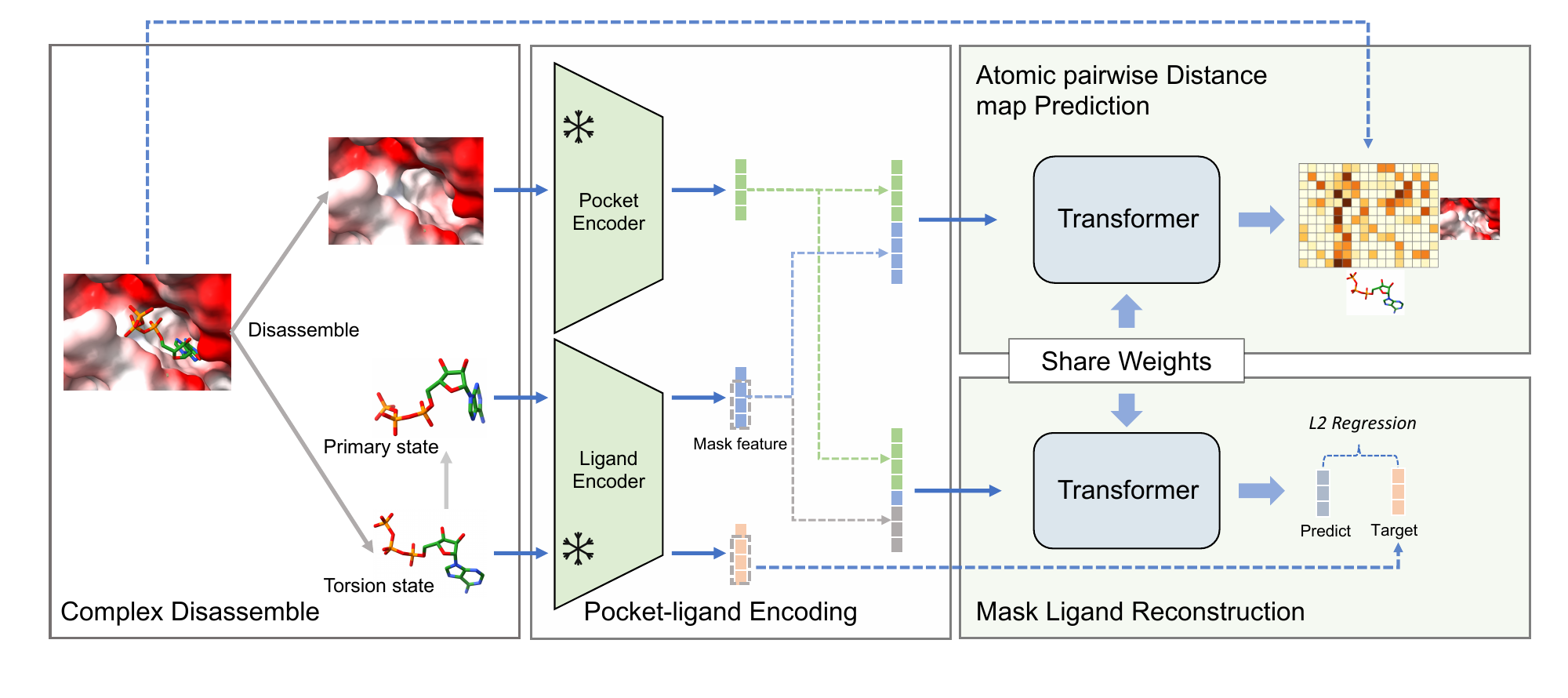}}
    \caption{Illustrations of BindNet consisting of three components. \textbf{Left}: To generate an input for the model, the protein-ligand complex is first disassembled, yielding the individual pocket and ligand structures. Then the primary ligand structure is approximated via RDKit and perturbation techniques, and this structure is treated as one of the inputs for the model. \textbf{Middle}: Pocket and ligand encoders are employed to extract their embeddings, with the weights remaining frozen during both training and testing phases. \textbf{Right}: Two novel pre-training objectives, i.e.~ADP and MLR, are introduced to learn binding representations.}
    \label{fig:illustrate}
    \end{center}
    \vskip -0.3in
\end{figure}

The architecture of BindNet is illustrated in Figure \ref{fig:illustrate}. Given a protein pocket and molecular ligand, two pre-trained encoders, designated as $\boldsymbol{\theta}_{P}$ and $\boldsymbol{\theta}_{L}$, are employed to secure preliminary representations for the pocket and ligand, denoted as $\mathbf{h}^{(0)}_{P}$ and $\mathbf{h}^{(0)}_{L}$, respectively. It is worth noting that the framework is versatile, and a variety of pre-existing encoders for pockets and ligands can be utilized. This paper uses the Uni-Mol encoders. It is important to understand that all pre-training and fine-tuning processes are conducted in the resulting representation space, while keeping the two pre-trained encoders fixed, in order to emphasize the learning of the subsequent interaction module. More precisely, the interaction module, represented by $\boldsymbol{\theta}_{I}$, is a N-layer 3D-invariant Transformer that takes both atom-wise and pairwise representations as input and generate pocket and ligand representations $\mathbf{h}^{(N)}_{P}$ and $\mathbf{h}^{(N)}_{L}$, as well as pairwise binding representations $\mathbf{h}^{(N)}_{PL}$. These representations will be subsequently used in the fine-tuning process to perform various downstream tasks.


\subsection{Pre-training Objectives}

To facilitate the pre-training process and accurately capture the intricate binding process between proteins and ligands, two objectives have been proposed. 


\subsubsection{Atomic Pairwise Distance Map Prediction}

According to previous biological and chemical studies~\citep{ponder2003force,alford2017rosetta}, the energy that arises from various non-bond interactions between proteins and ligands is closely associated with their inter-molecular distances. Therefore, several score functions~\citep{ballester2010machine} and deep learning methods~\citep{zhu2020binding, zheng2019onionnet, moon2022pignet} have utilized inter-molecular distance map to model the binding process, which has yielded significant improvements in binding related tasks.

Drawing inspiration from these findings, we propose utilizing the inter-molecular atom-wise distance map in a self-supervised manner to capture the intricate details of the interaction pattern. The inter-molecular distance matrix, denoted as $\mathcal{D}$, is initially derived from the original crystal structure of the complex. Each element, $d_{ij} \in \mathcal{D}$, represents the distance between the $i$-th atom in the pocket and the $j$-th atom in the ligand. Subsequently, we utilize the primary ligand $\mathcal{G}_{\hat{L}}$ and pocket $\mathcal{G}_P$ as inputs to predict the distance matrix $\mathcal{D}$ through regression. The specific objective function is:

\begin{footnotesize}
\begin{equation}
     \arg\min_{\boldsymbol{\theta}_I}\mathbb{E}_{(\mathcal{G}_{P}, \mathcal{G}_{\hat{L}})}\left[\mathcal{L}_{reg}\left(\textrm{MLP}\left(\mathbf{h}^{(N)}_{P\hat{L}}\right), \mathcal{D}\right)\right],
\end{equation}
\end{footnotesize}
where $\mathbf{h}^{(N)}_{P\hat{L}}$ represents the pair-wise embedding of $\mathcal{G}_{P}$ and $\mathcal{G}_{\hat{L}}$, $\mathcal{L}_{reg}$ denotes the $L_2$ regression loss.

Please note that our reason for using the primary ligand rather than the resultant ligand conformation in the complex data is to emphasize learning on the intricate interactions. Specifically, if we use the latter, this task focuses solely on learning the translation and rotation of the ligand to recover $\mathcal{D}$, thereby neglecting the crucial aspects of interaction information about inner changes in the ligand, such as variations in torsion angles when it binds to the target pocket.

\subsubsection{Mask Ligand Reconstruction}

While the aforementioned ADP objective measures the binding of the original protein and ligand to form the final complex structure, the following MLR objective is designed to reflect the conditional dependency relations between protein and ligand representations in the binding process. More precisely, we randomly mask the representation of the torsion ligand state and reconstruct it from the representation of the entire pocket and the remaining atoms of the primary ligand. 

To be specific, we replace the atom embeddings, denoted as $h_{i}$ in $\mathbf{h}^{(0)}_{\hat{L}}$ of the primary ligand with a learnable embedding $h_{m}$. Simultaneously, we mask the same corresponding set of atom embeddings in $\mathbf{h}^{(0)}_{L}$ of the torsion state ligand. These operations result in masked atom embeddings of torsion state ligand, as denoted as $\mathbf{h}^{(0)}_{L\mathbf{m}}$, which serves as the reconstruction target, along with the surrounding unmasked embeddings from the primary state ligand, denoted as $\mathbf{h}^{(0)}_{\hat{L} \backslash \mathbf{m}}$. Consequently, the objective of masked ligand reconstruction can be expressed as follows:

\begin{footnotesize}
\begin{equation}
    \arg\max_{\boldsymbol{\theta}_I}\mathbb{E}_{(\mathcal{G}_{P},\mathcal{G}_{\hat{L}})}\left[\text{P}\left(\mathbf{h}^{(0)}_{L\mathbf{m}} | \mathbf{h}^{(0)}_{\hat{L} \backslash \mathbf{m}},\mathbf{h}^{(0)}_{P}\right)\right] \\
     \simeq \arg\min_{\boldsymbol{\theta}_I}\mathbb{E}_{(\mathcal{G}_{P}, \mathcal{G}_{\hat{L}})}\left[\mathcal{L}_{reg}\left(\mathbf{h}^{(0)}_{L\mathbf{m}}, \tilde{\mathbf{h}}^{(0)}_{L\mathbf{m}}\right)\right],
\end{equation}
\end{footnotesize}

where we employ $L_2$ regression loss for reconstructing the target embeddings, $\tilde{\mathbf{h}}^{(0)}_{L\mathbf{m}}$ denotes the corresponding atom embeddings predicted by $\boldsymbol{\theta}_I$.


Our masking strategy differs from traditional approaches in two ways. Firstly, the proposed mask and reconstruction methodology is carried out in the representation space, rather than at the atom token level in most previous work~\citep{hu2019strategies, zhang2021motif, li2021effective}. This is because atom types are usually limited in molecular applications, which makes the task trivial and the model easy to fit as demonstrated in Mole-BERT~\citep{xia2022mole}. While conducting mask and reconstruction in the representation space helps to capture the intricate interaction patterns between pocket and ligand. These patterns are not solely dependent on ligand atom types but also involve atom positions and contextual information, which have already been well captured in pre-trained molecular representations. Secondly, the remaining primary ligand representation is used to recover the masked torsion molecular representation, which captures both the conditional dependencies between protein and molecular features and the change in conformation during the binding process.

\section{Experiments}
\label{headings}


BindNet is pre-trained on BioLip~\citep{yang2012biolip}, where we solely use the entries for regular ligands. For each complex, we extract the pocket-ligand segment by selecting residues within 8Å distance from the ligand as the pocket. The complex is removed from our pre-training dataset if no residues are present within 8Å distance from the ligand or if only hydrogen atoms are present in the ligand. Finally, we obtain a dataset with 458,252 pocket-ligand complexes. The model optimization is carried out using the Adam optimizer with a learning rate of 1e-4 and a weight decay of 1e-4. The mask ratio of MLR is set to 0.8, and ADP and MLR losses are treated equally. The model is trained for 30 epochs with a batch size of 32 batch size, which is completed on a machine equipped with 8-A100 GPUs. This section demonstrate our experiments on various binding related downstream tasks, including protein-ligand binding affinity prediction, virtual screening, and molecular docking.

\subsection{Protein-ligand Binding Affinity Prediction}

Protein-ligand binding affinity prediction seeks to anticipate the degree of interaction strength between proteins and ligands. We assess the performance of BindNet on two binding affinity prediction related tasks, LBA and LEP, as originally proposed in Atom3D~\citep{townshend2020atom3d}.

\subsubsection{Ligand Binding Affinity}

\paragraph{Data.}
Ligand Binding Affinity (LBA) is a regression task that involves predicting the binding affinity value. The protein-ligand complexes and their associated binding strengths are obtained from the PDBBind dataset~\citep{wang2005pdbbind}. The dataset is partitioned using a protein sequence identity threshold, resulting in two distinct splits: LBA 30\% (with a protein sequence identity threshold of 30\%) and LBA 60\% (with a protein sequence identity threshold of 60\%). We employ RMSE (Root Mean Square Error), Pearson correlation coefficient, and Spearman correlation coefficient, to evaluate BindNet. To ensure robustness of evaluation, we conduct three runs with different random seeds and report the mean values for the aforementioned metrics.
\paragraph{Baselines.}
We have compared BindNet with a diverse range of supervised methods, including sequence-based techniques such as DeepDTA, TAPE~\citep{rao2019evaluating}, and ProtTrans~\citep{elnaggar2021prottrans}; structure-based techniques, such as various variants of Atom3D, Holoprot~\citep{somnath2021multi}, and ProNet~\citep{wang2022learning}; as well as semi/self-supervised methods namely DeepAffinity, SMT-DTA, GeoSSL~\citep{liu2022molecular} and Uni-Mol.

\paragraph{Fine-Tuning BindNet.}
Since the LBA dataset has provided precise crystal structural information on the binding complex, we utilize the binding pocket and ligand from this complex structure as input to conduct the fine-tuning process. Specifically, we select the embeddings of the $\textrm{CLS}$ tokens from both pocket and ligand, which correspond to the first element of $\mathbf{h}^{(N)}_{P}$ and $\mathbf{h}^{(N)}_{L}$, respectively. These embeddings are concatenated and passed through a two-layer MLP to predict the binding affinity. The training process is supervised using a $L_2$ regression loss. Finally, we report the testing results based on the model that yields the best validation performance on the validation set.


\paragraph{Results.}


The experimental results are demonstrated in Table \ref{table:lba}. Comparing different deep learning approaches, we observe that structure-based methods generally outperform sequence-based methods. This finding is rooted in the rich interaction information intrinsic to structural data, which offers more detailed insights than sequences. Furthermore, comparing pre-training methods to deep learning methods, recent advancements in pre-training, such as GeoSSL and Uni-Mol, can significantly outperform deep learning methods. This demonstrates the effectiveness of self-supervised learning using large amounts of unlabeled data. Importantly, our proposed BindNet achieves the best results in terms of all metrics for both LBA 30\% and LBA 60\%, indicating the benefit to capturing knowledge from fine-grained interactions, as compared to learning individual protein or ligand representations. Notably, BindNet performs particularly well in LBA 30\%, which features strict data splitting and lower sequence identity between training and testing data. The substantial improvement underscores its superior generalizability by capturing essential interaction knowledge.

\begin{table}[ht]
    \centering
    \caption{Performance comparison of various methods on LBA dataset under different protein sequence identity split settings. The best and second-best results are highlighted in \textbf{bold} and \underline{underlined}, respectively (all tables below are presented in this format).}
    \label{table:lba}
    \scalebox{0.9}{
    \begin{tabular}{cccccccc}
        \toprule
        & & \multicolumn{3}{c}{LBA 30\%} & \multicolumn{3}{c}{LBA 60\%} \\
        \cmidrule(lr){3-5} \cmidrule(lr){6-8}
        Methods & Model & RMSE$\downarrow$ & Pearson$\uparrow$ & Spearman$\uparrow$ & RMSE$\downarrow$ & Pearson$\uparrow$ & Spearman$\uparrow$ \\
        \midrule
        \multirow{3}{*}{\shortstack{Sequence \\ based DL}} & DeepDTA & 1.866 & 0.472 & 0.471 & 1.762 & 0.666 & 0.663 \\
        & TAPE & 1.890 & 0.338 & 0.286 & 1.633 & 0.568 & 0.571 \\
        & ProtTrans & 1.544 & 0.438 & 0.434 & 1.641 & 0.595 & 0.588 \\
        \midrule
        \multirow{5}{*}{\shortstack{Structure \\ based DL}} & Atom3D-CNN & \underline{1.416} & 0.550 & 0.553 & 1.621 & 0.608 & 0.615 \\
        & Atom3D-ENN & 1.568 & 0.389 & 0.408 & 1.620 & 0.623 & 0.633 \\
        & Atom3D-GNN & 1.601 & 0.545 & 0.533 & 1.408 & 0.743 & 0.743 \\
        & Holoprot & 1.464 & 0.509 & 0.500 & 1.365 & 0.749 & 0.742 \\
        & ProNet & 1.463 & 0.551 & 0.551 & \underline{1.343} & \underline{0.765} & \underline{0.761} \\
        \midrule
        \multirow{4}{*}{\shortstack{Pre-training \\ Methods}} & DeepAffinity & 1.893 & 0.415 & 0.426 & - & - & - \\
        & SMT-DTA & 1.574 & 0.458 & 0.447 & 1.347 & 0.758 & 0.754 \\
        & GeoSSL & 1.451 & \underline{0.577} & \underline{0.572} & - & - & - \\
        & Uni-Mol & 1.434 & 0.565 & 0.540 & 1.357 & 0.753 & 0.750 \\
        & BindNet & \textbf{1.340} & \textbf{0.632} & \textbf{0.620} & \textbf{1.230} & \textbf{0.793} & \textbf{0.788} \\
        \bottomrule
    \end{tabular}
    }
\end{table}

\subsubsection{Ligand Efficacy Prediction}
\paragraph{Data.}
Ligand Efficacy Prediction (LEP) is a task to classify whether a ligand activates a specific protein when provided with both the active and inactive structural states. We follow the split defined in Atom3D based on the protein function. Typical classification measures such as the Area Under the Receiver Operating Characteristic (AUROC) and the Area Under the Precision-Recall Curve (AUPRC), are utilized as the evaluation metrics. Similar to LBA, we conduct three separate runs with varying random seeds and report the average results of the aforementioned metrics.

\paragraph{Baseline Methods.}
Our baseline methods include supervised techniques such as sequence-based approach DeepDTA and structure-based methods such as Atom3D-CNN, Atom3D-ENN, and Atom3D-GNN, along with pre-trained methods such as Uni-Mol and GeoSSL.

\paragraph{Fine-Tuning BindNet.}
We employ the pocket and ligand information provided by LEP as input, similar to LBA. Namely, we concatenate four pre-trained embeddings, incorporating the CLS tokens from the active structure's pocket and ligand, along with those of the inactive structure's pocket and ligand. Subsequently, these merged embeddings undergo a two-layer MLP for the final classification phase. The training process is supervised using cross-entropy.
Finally, we report the testing results based on the model that yields the best validation performance on the validation set.


\paragraph{Results.}
Table \ref{table:lep} presents our experimental results. BindNet exhibits superior performance compared to all supervised learning and pre-training methods, as measured by both AUROC and AUPRC metrics. Notably, the improvement over the second-ranking method (Uni-Mol) is substantial, with an AUROC improvement of 0.882 vs. 0.823 and an AUPRC improvement of 0.870 vs. 0.787. These significant deviations further validate the effectiveness of focusing on learning binding representations, rather than individual protein and ligand representations.


\vspace {-11pt}
\begin{table}[ht]
\centering
\caption{Comparison results on LEP datasets.}
\label{table:lep}
\scalebox{0.8}{
\begin{tabular}{cccc}
\hline
Methods & Model & AUROC$\uparrow$ & AUPRC$\uparrow$ \\
\hline
\multirow{1}{*}{\shortstack{Sequence based DL}} & DeepDTA & 0.696 & - \\
\cline{1-4}
\multirow{4}{*}{\shortstack{Structure \\ based DL}} & Atom3D-CNN & 0.589 & 0.483  \\
& Atom3D-ENN & 0.663 & 0.551 \\
& Atom3D-GNN & 0.681 & 0.598  \\
& GVP-GNN & 0.628  & - \\
\hline
\multirow{3}{*}{\shortstack{Pre-training \\ Methods}} & GeoSSL & 0.776 & 0.694 \\
& Uni-Mol & \underline{0.823} & \underline{0.787}\\
& BindNet & \textbf{0.882} & \textbf{0.870} \\
\hline
\end{tabular}
}
\end{table}

\subsection{Virtual Screening}
\paragraph{Data.}
DUD-E~\citep{mysinger2012directory} is a widely used benchmark for virtual screening, comprising 102 targets across multiple protein families. Each target contains an average of 224 active compounds and over 10,000 decoy compounds. We employ a 3-fold cross-validation for training and evaluation, and our dataset split setting is consistent with AttentionDTI~\citep{yazdani2022attentionsitedti} and DrugVQA~\citep{zheng2020predicting}, ensuring that similar targets are kept within the same fold to facilitate a fair comparison. Several widely used measures on DUD-E are employed in our evaluation, including AUROC and the ROC Enrichment metric (denoted as RE). 


\paragraph{Baseline Methods.}

Various baseline methods are utilized in our experiments, including docking programs AutoDock Vina~\citep{trott2010autodock} and Smina~\citep{koes2013lessons}, traditional statistical machine learning methods such as RF-score and NNScore~\citep{durrant2011nnscore}, deep learning methods such as 3D-CNN, Graph-CNN~\citep{torng2019graph}, AttentionSiteDTI, DrugVQA, as well as pre-training methods such as Uni-Mol and CoSP.

\vspace {-11pt}
\begin{table}[htbp]
\centering
\caption{Performance comparison of different methods on DUD-E.}
\label{table:dude}
\scalebox{0.8}{
\begin{tabular}{ccccccc}
\hline
Methods & Model & AUC$\uparrow$ & 0.5\% RE$\uparrow$ & 1.0\% RE$\uparrow$ & 2.0\% RE$\uparrow$ & 5.0\% RE$\uparrow$ \\ \hline
\multirow{2}{*}{\shortstack{Docking \\ based}} & AutoDock Vina & 0.716 & 9.139 & 7.321 & 5.811 & 4.444 \\
& Smina & 0.696 & - & - & - & - \\
\hline
\multirow{2}{*}{\shortstack{Scoring function \\ based ML}} & RF-score & 0.622 & 5.628 & 4.274 & 3.499 & 2.678 \\ 
& NNScore & 0.584 & 4.166 & 2.980 & 2.460 & 1.891 \\
\hline
\multirow{4}{*}{\shortstack{Supervised \\ based DL}} & 3D-CNN & 0.868 & 42.559 & 26.655 & 19.363 & 10.710 \\
& Graph-CNN & 0.886 & 44.406 & 29.748 & 19.408 & 10.735 \\
& DrugVQA & \textbf{0.972} & 88.170 & 58.710 & 35.060 & \textbf{17.390} \\ 
& AttentionSiteDTI & \underline{0.971} & \underline{101.740} & \underline{59.920} & \underline{35.070} & \underline{16.740} \\
\hline
\multirow{3}{*}{\shortstack{Pre-training \\ Methods}} & CoSP & 0.901 & 51.048 & 35.978 & 23.681 & 12.212 \\ 
& Uni-Mol & 0.945 & 82.586 & 50.206 & 30.162 & 14.789 \\ 
& BindNet & 0.960 & \textbf{105.277} & \textbf{61.602} & \textbf{35.150} & 16.185\\ 
\hline
\end{tabular}
}
\end{table}

\paragraph{Fine-Tuning BindNet.}
For each target, we extract residues within 6 Å distance from the crystal ligand as the pocket. We utilize cross-entropy based on the output of a MLP, with the CLS token embeddings of the pocket and ligand serving as inputs. Due to the significant imbalance between negative pairs (comprising inactive compounds) and positive pairs (comprising active compounds), we dynamically adjust the sampling weights to ensure that each batch contains an equal number of negative and positive samples. We report the mean performance of 3-fold cross-validation.


\paragraph{Results.}
As demonstrated in Table \ref{table:dude}, BindNet achieves the highest performance with respect to three RE metrics and delivers competitive results in terms of the AUC score and the other RE metric. In particular, when compared to the other self-supervised learning methods such as Uni-Mol and CoSP, BindNet consistently outperforms them by a significant margin across all evaluation metrics. However, some supervised learning methods, such as DrugVQA and AttentionSiteDTI, display more stable and outstanding results. This phenomenon may be caused by the decoy bias hidden in the DUD-E dataset, as discussed in ~\cite{gonczarek2016learning} and ~\cite{chen2019hidden}. Specifically, the selection criteria for inactive compounds within DUD-E involve choosing compounds with similar physical properties to active compounds but differing topological structures, which may introduce a bias that emphasizes discrepancies between active and inactive compounds rather than focusing on protein-ligand interactions. Consequently, supervised learning methods may be more advantageous as they can directly incorporate and accommodate such biases. 

To validate our assumption, we conduct further experiments on the AD dataset~\citep{chen2019hidden}, which improves upon DUD-E by employing active compounds from other targets as decoys for the current target and effectively mitigates decoy bias. We evaluate the top four methods from the DUD-E dataset, and find a significantly decrease in performance shown in Table \ref{table:ad}, especially for supervised learning methods like DrugVQA and AttentionSiteDTI. This supports our claim that supervised methods tend to capture data biases rather than authentic interaction information. Conversely, Uni-Mol and BindNet perform notably better, with BindNet significantly surpassing Uni-Mol, reaffirming the criticality and benefit of learning intricate interaction patterns.


\vspace {-11pt}
\begin{table}[htbp]
\centering
\caption{Performance comparison on AD dataset.}
\label{table:ad}
\scalebox{0.9}{
\begin{tabular}{cccccc}
\hline
Methods & AUC$\uparrow$ & 0.5\% RE$\uparrow$ & 1.0\% RE$\uparrow$ & 2.0\% RE$\uparrow$ & 5.0\% RE$\uparrow$ \\ \hline
DrugVQA & 0.48 & 3.00 & 2.44 & 2.15 & 1.79 \\
AttentionSiteDTI & 0.47 & 2.79 & 2.32 & 2.24 & 1.63 \\
\hline
Uni-Mol & \underline{0.56} & \underline{4.92} & \underline{3.70} & \underline{2.82} & \underline{2.00} \\
BindNet & \textbf{0.64} & \textbf{6.98} & \textbf{4.45} & \textbf{3.21} & \textbf{2.97} \\ \hline
\end{tabular}
}
\label{table:results}
\end{table}




\vspace{-5pt}
\subsection{Molecular Docking}
\paragraph{Data.}
We employ the same dataset as Uni-Mol to evaluate BindNet's performance in protein-ligand docking. The dataset includes the general set from PDBbind v2020 as the training set, with CASF-2016 as the test set, with data overlap removed for fair comparison. BindNet has learned the structural complex information in the pre-training stage, hence, we retrain it with the overlap removed data. Our evaluation metric involves calculating the Root Mean Square Deviation (RMSD) between predicted and actual positions, presenting the percentage of values falling below a threshold.

\paragraph{Baselines.}
Our baselines consist of a pre-training method Uni-Mol and various score-based methods, including AutoDock Vina, Vinardo~\citep{quiroga2016vinardo}, Smina, and AutoDock4~\citep{morris2009autodock4}. These results are obtained directly from the original Uni-Mol paper.

\paragraph{Fine-Tuning BindNet.}
During the fine-tuning process, we initially predict the distance map of the docking complex, by using the pairwise representations $\mathbf{h}^{(N)}_{PL}$. Subsequently, we employ a post-processing approach to acquire the docked pose, by using gradient descent optimization, which is a common technique utilized in several typical docking methods \citep{lu2022tankbind}.


\paragraph{Results.}

\vspace {-11pt}
\begin{table}[htbp]
\centering
\caption{Performance comparison on docking pose prediction.}
\label{table:dock}
\scalebox{0.9}{
\begin{tabular}{cccccc}
\hline
Methods & 1.0 Å$\uparrow$ & 1.5 Å$\uparrow$ & 2.0 Å$\uparrow$ & 3.0 Å$\uparrow$ & 5.0 Å$\uparrow$ \\
\hline
Autodock Vina & 44.21 & 57.54 & 64.56 & 73.68 & 84.56 \\
Vinardo & 41.75 & 57.54 & 62.81 & 69.82 & 76.84 \\
Smina & \textbf{47.37} & 59.65 & \underline{65.26} & 74.39 & 82.11 \\
Autodock4 & 21.75 & 31.58 & 35.44 & 47.02 & 64.56 \\
\hline
Uni-Mol & 43.16 & \underline{68.42} & \textbf{80.35} & \underline{87.02} & \underline{94.04} \\
BindNet & \underline{45.26} & \textbf{69.82} & \textbf{80.35} & \textbf{89.12} & \textbf{94.38} \\
\hline
\end{tabular}
}
\end{table}

As depicted in Table~\ref{table:dock}, BindNet surpasses all other baseline methods. While Uni-Mol also outperforms all other score-based methods, it struggles to perform as effectively in the setting at the percentage under the 1.0Å threshold, compared to Smina and Autodock Vina. However, BindNet effectively manages to combat this issue by enhancing the percentage of results from Uni-Mol's 43.16\% to BindNet's 45.26\%. This indicates that BindNet's representations capture more precise interactions between ligands and proteins, leading to more accurate docking poses.

\section{Ablation Study}

\subsection{Effects of varying pre-training objectives}

\begin{figure}[ht]
    \vskip -0.1in
    \begin{center}
    \scalebox{0.8}{
    \centerline{\includegraphics[width=15cm]{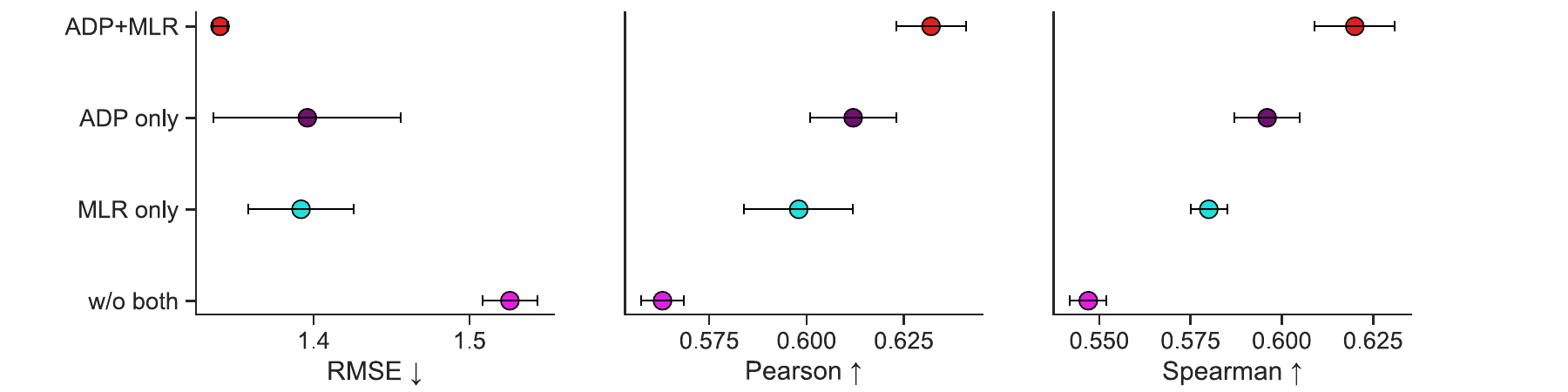}}
    }
    \caption{\label{figure:abi}Ablation study on pre-training objectives, where the scatter represents the mean value of each metric and the error bar shows the standard deviation.}
    \end{center}
    \vskip -0.2in
\end{figure}

As BindNet incorporates two pre-training objectives, we conduct an ablation study to validate the impact of each loss using the LBA dataset, shown in Figure ~\ref{figure:abi}. It is evident that both ADP and MLR play crucial roles in BindNet's performance. The model trained solely on ADP or MLR outperforms the one without pre-training. Moreover, the best results are obtained by combining both strategies and training the model in a multi-task manner, demonstrating the complementary nature of these two strategies: ADP focuses on extracting binding knowledge from complexes, while MLR learns how to construct a ligand to form a stable binding pattern given a pocket target. The combination of these strategies results in a more comprehensive and robust interaction aware representation.

\subsection{Out-of-Distribution Evaluation on Varying Complex Structure}
\label{Precise crystal structure influence}

While 3D structure based deep learning methods often generate prediction results that are comparable to those of pre-training methods, they are usually evaluated in the in-distribution settings. In LBA, LEP and DUD-E, crystal structure, docking conformations and RDKit generated structures are used for both training and testing, respectively. As a result, it is unclear whether these deep learning models are capable of out-of-distribution generalization, a crucial issue in machine learning, particularly for robust and trustworthy learning in scientific fields.


We evaluate out-of-distribution capability with three new settings in LBA, with varying complex structure: CD (crystal conformations for training and docking complex conformations for testing), DD (docking complex conformations for training and testing), and RR (RDKit generated conformations for training and testing). The original setting CC uses crystal complex conformations for both training and testing. We utilize Atom3D-CNN and Atom3D-GNN as representative examples of 3D structure based supervised learning methods, exhibiting high performance on LBA (Table \ref{table:lba}). As for pre-training methods, we evaluate Uni-Mol and BinNet. It should be noted that Atom3D-CNN and Atom3D-GNN are not evaluated in the RR setting as they only accept complex structure as input.

\begin{figure}[ht]
    \vskip -0.1in
    \begin{center}
    \scalebox{0.8}{
    \centerline{\includegraphics[width=15cm]{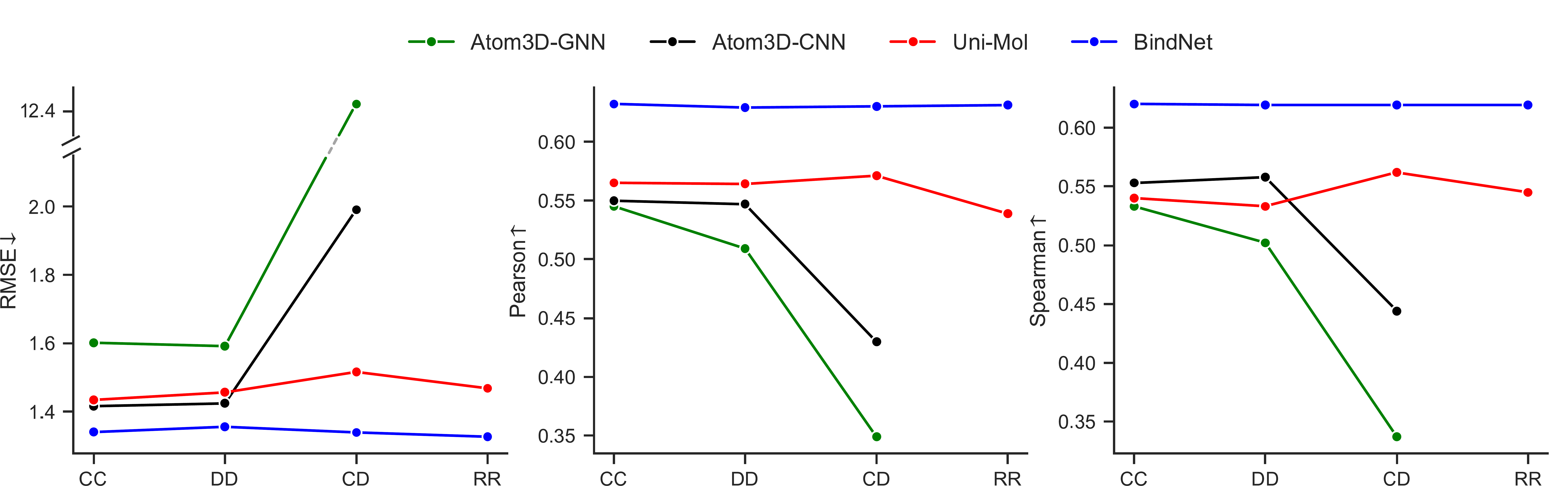}}
    }
    \caption{\label{figure:abi_2}Performance of various methods across different settings with varying complex structure.}
    \end{center}
    \vskip -0.1in
\end{figure}

Figure~\ref{figure:abi_2} shows that pre-training methods exhibit superior out-of-domain generalization ability than supervised learning methods. Specifically, both Atom3D-CNN and Atom3D-GNN perform well under in-distribution settings CC and DD, but their out-of-distribution performance decreases significantly, as evidenced by the increase from 1.601 to 12.475 in RMSE when transitioning from CC to CD for Atom3D-GNN. Hence, one can infer that these 3D structure based supervised learning methods merely learn a data-fitting function, without truly capturing protein-ligand interaction patterns. Conversely, Uni-Mol and BindNet perform consistently well across all settings, even with randomly initialized conformations as input, demonstrating pre-training approach's efficacy in acquiring intrinsic data representations. Moreover, BindNet consistently outperforms Uni-Mol in both in-distribution and out-of-distribution evaluations, emphasizing the superiority of learning interaction-aware representations over individual protein and ligand representations.

\vspace{-5pt}
\section{Conclusion}
\vspace{-5pt}
This paper proposed a novel self-supervised pre-training method called BindNet for the purpose of learning protein-ligand binding representations. Unlike previous pre-training approaches that focus on individual protein and ligand representations, BindNet places greater emphasize on learning the binding representations using a Transformer-based interaction module, with fixed protein and ligand encoders as input. We proposed two new objective functions, i.e.~ADP and MLR, to facilitate the pre-training from fine-grained interaction signals. Our analysis indicate that these objectives are crucial for learning comprehensive and robust interaction aware representations, as they play complementary roles. By applying BindNet to various downstream binding related tasks, such as protein-ligand binding affinity prediction, virtual screening, and protein-ligand docking, we demonstrate that our approach significantly outperforms existing supervised and pre-training methods. Besides, our ablation study show that BindNet successfully learns meaningful, robust representations that are capable of dealing with varying complex structures in out-of-distribution settings.

Although our primary focus is on the protein-ligand binding domain, the BindNet framework, as a powerful tool for learning binding representations, has great potential for extension to other bio-related binding tasks, such as protein-protein interactions and antigen-antibody recognition.



\bibliography{BindNet}
\bibliographystyle{iclr2024_conference}


\end{document}